\documentclass[]{llncs}
\usepackage[T1]{fontenc}
\usepackage{graphicx}
\usepackage{pifont}
\usepackage{booktabs}
\usepackage{amsmath}
\usepackage{url}
\usepackage{listings}
\begin{document}
\title{Overview of SCIDOCA 2025 Shared Task on Citation Prediction, Discovery, and Placement}
%
%
\author{An Dao\inst{1}\and
Vu Tran\inst{2}\and
Le-Minh Nguyen\inst{3}\and
Yuji Matsumoto\inst{1}
}
\authorrunning{}
%
\institute{RIKEN Center for Advanced Intelligence Project (AIP), Japan\inst{1}\\
The Institute of Statistical Mathematics, Japan\inst{2}\\
Japan Advanced Institute of Science and Technology\inst{3}\\
\email{}
}
\maketitle              
\begin{abstract}
We present an overview of the SCIDOCA 2025 Shared Task, which focuses on citation discovery and prediction in scientific documents. 
The task is divided into three subtasks: (1) Citation Discovery, where systems must identify relevant references for a given paragraph; (2) Masked Citation Prediction, which requires selecting the correct citation for masked citation slots; and (3) Citation Sentence Prediction, where systems must determine the correct reference for each cited sentence. 
We release a large-scale dataset constructed from the Semantic Scholar Open Research Corpus (S2ORC), containing over 60,000 annotated paragraphs and a curated reference set. 
The test set consists of 1,000 paragraphs from distinct papers, each annotated with ground-truth citations and distractor candidates. 
A total of seven teams registered, with three submitting results. 
We report performance metrics across all subtasks and analyze the effectiveness of submitted systems. 
This shared task provides a new benchmark for evaluating citation modeling and encourages future research in scientific document understanding.
The dataset and task materials are publicly available at \url{https://github.com/daotuanan/scidoca2025-shared-task}.

\end{abstract}

\section{Introduction}
Citation recommendation or prediction—the task of identifying relevant references for a given scientific text—is a fundamental problem in bibliometrics and academic research. 
Effective citation recommendation systems assist scholars in discovering pertinent literature, support publishers in improving academic search and indexing services (e.g., Google Scholar, Scopus), and contribute to analyzing the influence and impact of scientific work. 
Beyond merely linking papers, understanding citation contexts—the textual fragments surrounding citations—enables more nuanced insights into citation intent, influence, and knowledge propagation.

Within the broad domain of citation analysis, related research topics include citation count prediction \cite{yan2011citation,pobiedina2016citation,li2019neural,ma2021deep}, citation recommendation \cite{ali2020deep,farber2020citation,ali2021overview}, and citation context analysis \cite{small1980co,hernandez2016survey,anderson2023citation}. 
This shared task focuses on citation context analysis, which involves identifying citations relevant to a specific textual context and understanding the semantics behind citation usage. 
Accurate citation context analysis can further improve downstream applications such as scientific summarization, knowledge graph construction, and impact evaluation.

Several prior shared tasks have explored problems related to scientific document understanding. 
For instance, the CL-SciSumm 2016 Shared Task \cite{jaidka2016overview} addressed scientific document summarization by identifying and summarizing cited content within full-text papers. 
Similarly, various citation recommendation datasets such as CiteSeerX~\cite{wu2015citeseerx}, RefSeer~\cite{huang2014refseer}, ACL-AAN~\cite{radev2013acl}, ACL-ARC~\cite{bird2008acl}, arXiv CS\cite{farber2018high}and unarXiv~\cite{saier2019bibliometric}  been used to develop and evaluate models for recommending references based on citation contexts. 
While these efforts have significantly advanced the field, they typically rely on full-text access, including the body, figures, and complete reference sections. 

To address the practical challenge of limited access to full-text articles, the SCIDOCA 2025 Shared Task centers on citation discovery and prediction using only partial textual information—specifically, the title and abstract of candidate papers. 
This reflects real-world scenarios frequently encountered in open-access repositories, preprint servers, and web-based academic search engines, where only metadata and abstracts are available.
The shared task is composed of three subtasks that together aim to improve citation modeling under such constrained conditions.
The first subtask, Citation Discovery, requires systems to identify relevant references for a given scientific paragraph from a list of candidate papers. 
This simulates citation recommendation in a setting where only limited metadata is available for candidate references. 
The second subtask, Masked Citation Prediction, challenges systems to recover specific citations that have been masked in a paragraph. 
Each placeholder must be accurately filled with the correct citation, mimicking tasks such as repairing incomplete manuscripts or automatically enhancing draft documents.
The third subtask, Citation Sentence Prediction, focuses on sentence-level citation placement.
Here, systems must determine for each sentence whether a citation is needed and, if so, identify the most appropriate one. 
This fine-grained task is particularly relevant for supporting scientific writing and editing tools.
By modeling citation behavior using limited resources, SCIDOCA 2025 promotes the development of practical and generalizable tools for real-world scholarly communication tasks.

To ensure fairness and replicability, we restrict participants to use only the data provided by the organizers, disallowing any external data sources. This constraint emphasizes building robust models under limited resource conditions and prevents data leakage or unfair advantages stemming from pretrained citation-specific datasets.

The dataset for this shared task is derived from the Semantic Scholar Open Research Corpus (S2ORC) \cite{lo2020s2orc}, one of the largest publicly available corpora of scientific papers. S2ORC contains structured full text for over 8 million open-access papers, metadata, and citation graphs across multiple scientific domains, making it well-suited for building citation-related tasks.

In summary, this shared task aims to advance the development of citation discovery and prediction models that operate effectively with limited textual information, fostering progress towards practical citation recommendation systems applicable to real-world, open-access scientific search scenarios.

\section{Task Description}
\subsection{Overview}
The SCIDOCA 2025 Shared Task focuses on Citation Prediction, Discovery, and Placement within scientific documents. Participants will be challenged to develop models capable of handling various citation-related tasks, including:

\begin{itemize}
    \item \textbf{Subtask 1: Citation Discovery:} Given a scientific paragraph and a list of candidate references, predict which citations are relevant to the paragraph without specifying their exact placement.

    \item \textbf{Subtask 2: Masked Citation Prediction:} Recover the correct citation for each masked citation slot (e.g., \texttt{[MASK1]}) in a paragraph, using the surrounding context and a set of candidate references.

    \item \textbf{Subtask 3: Citation Sentence Prediction:} For each sentence in a scientific paragraph, predict the appropriate citation(s) if any, identifying where citations should be placed for proper attribution.
\end{itemize}

The shared task is designed to assess models' abilities to understand intricate citation networks in scientific discourse while also evaluating their effectiveness in handling domain-specific knowledge. Participants will work with a curated dataset of scientific papers across multiple disciplines to ensure a comprehensive evaluation of their approaches.

\subsection{Motivation and Impact}
The SCIDOCA 2025 Shared Task aims to address the growing need for automated citation systems that assist researchers in managing the ever-expanding corpus of scientific literature. By improving citation discovery and placement, this task could lead to advancements in:

\begin{itemize}
    \item \textbf{Efficient Literature Review:} Helping researchers quickly find relevant work and ensuring proper attribution of ideas.
    \item \textbf{Improved Scientific Writing Tools:} Automating citation insertion to enhance the drafting process and reduce manual effort.
    \item \textbf{Citation Network Analysis:} Enabling better understanding of citation behaviors across scientific domains and improving citation-based metrics.
    \item \textbf{Domain-Specific Adaptation:} Evaluating how well models generalize across different scientific disciplines and citation styles.
\end{itemize}

\subsection{Subtask 1: Citation Discovery}
\textbf{Objective:} Predict relevant citations for a paragraph without specifying the exact sentence where the citation belongs.

\textbf{Input:}
\begin{itemize}
    \item \textbf{Paragraph:} A text passage from a scientific document that doesn’t contain citations.
    \item \textbf{Candidate References:} A list of potential references, which includes both the correct citations and distractors (irrelevant but plausible citations).
\end{itemize}

\textbf{Output:}
The system should return a list of citations from the candidate references that it deems relevant to the input paragraph. These are the references that support the claims or ideas in the paragraph, without requiring the system to specify where exactly they should be placed.

\textbf{Example Input:}
\begin{lstlisting}
{
  "paragraph": "Recent advances in natural language processing have significantly improved the performance of models on various tasks such as machine translation and question answering.",
  "candidate_references": [
    "[Vaswani et al. 2017]",
    "[Devlin et al. 2019]",
    "[Brown et al. 2020]",
    "[Radford et al. 2018]"
  ]
}
\end{lstlisting}

\textbf{Output:}
\begin{lstlisting}
{
  "predicted_citations": [
    "[Vaswani et al. 2017]",
    "[Devlin et al. 2019]"
  ]
}
\end{lstlisting}

\textbf{Evaluation:}
For each paragraph \(i\), let \(TP_i\), \(FP_i\), and \(FN_i\) denote true positives, false positives, and false negatives, respectively, based on the predicted and gold citation sets.

\begin{itemize}
    \item \textbf{Precision} for paragraph \(i\):
    \[
    Precision_i = \frac{TP_i}{TP_i + FP_i}
    \]
    Measures the proportion of correctly predicted citations among all predictions, reflecting relevance.
    
    \item \textbf{Recall} for paragraph \(i\):
    \[
    Recall_i = \frac{TP_i}{TP_i + FN_i}
    \]
    Measures the proportion of correctly predicted citations among all gold citations, reflecting completeness.
    
    \item \textbf{F1-Score} for paragraph \(i\):
    \[
    F1_i = 2 \times \frac{Precision_i \times Recall_i}{Precision_i + Recall_i}
    \]
\end{itemize}

\textbf{Evaluation across the dataset:} The final metric is the weighted average of paragraph-level F1 scores, weighted by the number of gold citations \(GT_i\) in each paragraph:

\[
F1 = \frac{\sum_i GT_i \times F1_i}{\sum_i GT_i}
\]

\subsection{Subtask 2: Masked Citation Prediction}
\textbf{Objective:} Predict the correct citation for each masked citation slot within a paragraph where the citation has been removed.

\textbf{Input:}
\begin{itemize}
    \item \textbf{Paragraph:} A paragraph where one or more citation slots have been masked (replaced by a placeholder such as [MASK1], [MASK2], etc.).
    \item \textbf{Candidate References:} A list of potential references, including both correct citations and distractors.
\end{itemize}

\textbf{Output:}
The system should return a mapping between each masked slot (e.g., [MASK1], [MASK2]) and the predicted citation from the candidate references that best fits the original context of that slot.



\textbf{Example Input:}
\begin{lstlisting}
{
   "paragraph": "Transformer models like BERT [MASK1] and GPT-3 [MASK2] have revolutionized natural language processing tasks. These models [MASK3] continue to set benchmarks across various domains.",
   "candidate_references": [
      "[Vaswani et al. 2017]",
      "[Devlin et al. 2019]",
      "[Brown et al. 2020]",
      "[Radford et al. 2018]"
   ]
}
\end{lstlisting}

\textbf{Output:}
\begin{lstlisting}
{
   "predicted_citations": {
      "[MASK1]": "[Devlin et al. 2019]",
      "[MASK2]": "[Brown et al. 2020]",
      "[MASK3]": "[Radford et al. 2018]"
   }
}
\end{lstlisting}

\textbf{Evaluation:}
Let \(M_i\) be the number of masked citation positions in paragraph \(i\). Evaluation is done per masked position:
    \textbf{Macro-Averaged F1}: Compute F1 for each citation label (masked slot) and average across all labels.
\textbf{Evaluation across dataset:} Metrics are aggregated over all masked citations and weighted by the number of masks \(M_i\) in each paragraph.

\subsection{Subtask 3: Citation Sentence Prediction}
\textbf{Objective:} Given a paragraph, predict the correct citation for each sentence that contains a citation.

\textbf{Input:}
\begin{itemize}
    \item \textbf{Paragraph:} A multi-sentence paragraph without any explicit citation markers.
    \item \textbf{Candidate References:} A list of potential citations, including both correct citations and distractors.
\end{itemize}

\textbf{Output:}
The system should return a list of sentence-level citation predictions. For each sentence in the paragraph, the system must indicate whether a citation is required, and if so, predict one or more relevant citations from the candidate reference list. Sentences that do not require a citation should return an empty list.

\textbf{Example Input:}
\begin{lstlisting}
{
  "paragraph": [
    "Transformer models have transformed the field of NLP.",
    "One of the most influential models is BERT.",
    "We will investigate the results of BERT models.",
    "GPT-3 has further pushed the boundaries of language modeling."
  ],
  "candidate_references": [
    "[Vaswani et al. 2017]",
    "[Devlin et al. 2019]",
    "[Brown et al. 2020]",
    "[Radford et al. 2018]"
  ]
}
\end{lstlisting}

\textbf{Output:}
\begin{lstlisting}
{
  "sentence_citations": [
    {
      "sentence": "Transformer models have transformed the field of NLP.",
      "predicted_citation": ["[Vaswani et al. 2017]"]
    },
    {
      "sentence": "One of the most influential models is BERT.",
      "predicted_citation": ["[Devlin et al. 2019]"]
    },
    {
      "sentence": "We will investigate the results of BERT models.",
      "predicted_citation": ["[empty]"]
    },
    {
      "sentence": "GPT-3 has further pushed the boundaries of language modeling.",
      "predicted_citation": ["[Brown et al. 2020]"]
    }
  ]
}
\end{lstlisting}

\textbf{Evaluation:}
Two sentence types exist:

\begin{itemize}
    \item \textbf{Sentences with citations:} Evaluate predicted citations against gold citations using sentence-level precision, recall, and F1.
    
    \item \textbf{Sentences without citations:} Evaluate whether the system correctly predicts no citations. This is measured via accuracy on empty predictions.
\end{itemize}

\textbf{Metrics:}

\begin{itemize}
    \item \textbf{Sentence-Level F1}: Computed by comparing predicted citations with gold citations for each sentence containing citations.
    
    \item \textbf{Micro F1}: This metric aggregates contributions from all sentences to compute an overall F1 score, providing a balanced measure of precision and recall across the dataset.    
    
    \item \textbf{Accuracy for No-Citation Sentences}: Proportion of sentences without gold citations correctly predicted as having no citations.
\end{itemize}

$\text{Accuracy}{\text{no-citation}} = \frac{1}{|S{nc}|} \sum_{i \in S_{nc}} I(\hat{y}_i = \emptyset)$

\textbf{Evaluation considerations:} Given the high prevalence of sentences without citations, relying solely on F1 could reward models that naively predict no citations for all sentences. Therefore, we consider both F1 and no-citation accuracy to provide a balanced evaluation across models.

\subsection{Data Usage Rules Summary}
\begin{itemize}
    \item \textbf{No External Data Transmission:} Systems must operate offline and cannot send any provided data (training or test) to external services or APIs.
    \item \textbf{No Human Intervention:} Systems must function autonomously during test-time inference, with no manual adjustments or parameter tuning.
    \item \textbf{Restricted Use of Non-Organizer Data:}
    \begin{itemize}
        \item External citation-related datasets or services (e.g., CrossRef, PubMed) are prohibited.
        \item General-purpose pretrained models (e.g., BERT) are allowed if unrelated to citations.
        \item Citation-related pretrained models (e.g., SPECTER, Galactica) are prohibited.
    \end{itemize}
\end{itemize}

\section{Dataset Construction}

\subsection{Overview}
The dataset for the shared task is carefully constructed to ensure diversity in citation contexts, broad domain coverage, and a well-structured dataset split. This section details the sources, annotation process, dataset structure, and quality control measures.

\subsection{Data Sources}
The dataset for this shared task is constructed using the Semantic Scholar Open Research Corpus (S2ORC) \cite{lo2020s2orc}. 
S2ORC is a large-scale dataset of academic papers, containing 81.1 million English-language papers across multiple domains. 
Among these, 8.1 million papers have structured full-text data, making them particularly valuable for citation-based tasks. 
The entire dataset spans 449GB and covers a wide range of scientific disciplines.

To construct the dataset, we extract structured citation information from S2ORC, ensuring high-quality data for the shared task. 
Preprocessing steps are applied to clean the text, filter incomplete references, and align citations with their corresponding textual contexts.

\subsection{Data Utilization}
S2ORC is a large-scale dataset spanning multiple scientific domains, providing structured metadata and full-text content for academic papers. For the shared task, we extract and utilize the following components:

\paragraph{Abstract} Used to understand the high-level summary of a paper, which may provide contextual information for citation prediction.

\paragraph{Authors} Metadata about the authors can help in citation attribution and understanding citation behaviors.

\paragraph{Body Text} The main content of the paper, from which we extract citation contexts for different subtasks.

\paragraph{References} The reference list provides the ground-truth citations for each paper. We use this to identify which citations are relevant for a given paragraph.

By leveraging these components, we construct a dataset suitable for citation discovery, masked citation prediction, and citation sentence prediction. Preprocessing steps ensure that the extracted text and references are correctly aligned and formatted for each subtask.

\subsection{Filtering Process}
To construct a high-quality dataset for the shared task, we apply a structured filtering process to extract relevant papers and references from S2ORC. The dataset is divided into the following subsets:

\paragraph{Primary Paper Set}  
This set contains papers that serve as the primary input for citation discovery and prediction tasks. Each paper includes:
- Abstract  
- Full-text  
- Reference links that must be present in the Curated Reference Set.  

\paragraph{Extended Primary Paper Set}  
A broader version of the Primary Paper Set, including additional papers with more reference coverage. Papers in this set include:
- Abstract  
- Full-text  
- Reference links that must be present in the Comprehensive Reference Set.  

\paragraph{Curated Reference Set}  
This set consists of 10,154,823 reference papers, selected for their relevance to the primary papers. Each reference includes:  
- Abstract  
- Full-text  
- No restrictions on reference links.  

\paragraph{Comprehensive Reference Set}  
A larger reference set containing 75,555,240 papers with broader coverage. Each reference includes:  
- Abstract  
- No restrictions on reference links.  

By organizing the dataset into these structured subsets, we ensure a well-defined relationship between input papers and their references, facilitating effective evaluation in the shared task.

\subsection{Annotation Process}

The annotation process for the SCIDOCA 2025 Shared Task is fully automated, leveraging citation metadata available in the S2ORC (Semantic Scholar Open Research Corpus). Specifically, we utilize the original citation spans, citation identifiers, and positional information embedded in the body text of the source papers.

Each paragraph is annotated with its corresponding ground-truth citations based on citation spans that explicitly reference cited works within the paragraph. This method ensures accurate and consistent labeling without the need for manual verification.

To create a realistic and challenging evaluation setup, we introduce distractor citations. These are drawn from other citations made by the same paper but located in different paragraphs. This strategy guarantees that the distractors are topically relevant and contextually plausible, thereby increasing the difficulty of the prediction task.

\subsection{Dataset Splits}

The dataset for the SCIDOCA 2025 Shared Task is split into several subsets, providing train and test data for the different subtasks. 

\subsubsection{Train}

The training dataset for the SCIDOCA 2025 Shared Task comprises 10,000 full-text scientific papers, carefully selected to ensure a diverse and citation-rich corpus. These documents serve as the foundation for generating citation contexts and candidate references across all three subtasks.
In total, the dataset contains 61,556 paragraphs, with an average of 5.62 sentences per paragraph. Each paragraph has, on average, 2.11 gold citations and 18.04 candidate references. Paragraphs vary in length from 1 to 86 sentences, and the number of gold citations per paragraph ranges from 1 to 36. The number of citation candidates per paragraph ranges from 0 to 151, reflecting the diversity and complexity of the citation prediction task.

\subsubsection{Test}

The test set consists of 1,000 paragraphs, each selected from a distinct scientific paper. This design ensures that no paper contributes more than one paragraph to the test set. By doing so, we prevent models from leveraging additional paragraphs of the same paper to eliminate distractors or make inferences beyond the provided context.
Each test paragraph is annotated with ground-truth citations based on original citation metadata in the S2ORC dataset. Distractor citations are sourced from other references within the same paper but outside the target paragraph, ensuring that all candidates are plausible and topically related.
This setup aims to fairly evaluate a model’s ability to understand a single paragraph in isolation and to assess its performance across all three subtasks without external contextual leakage.

\subsection{Data Format and Structure}
The dataset is provided in a structured format (e.g., JSON). Below is an example of the format for each subtask:

\begin{itemize}
    \item \textbf{Citation Discovery:} Each entry consists of a paragraph and a list of candidate references.
    \item \textbf{Masked Citation Prediction:} Each paragraph contains masked citation slots, and the task is to predict the correct citations.
    \item \textbf{Citation Sentence Prediction:} Each sentence in a paragraph must be assigned the correct citation(s).
\end{itemize}

\section{Team Methodologies}

\subsection{Team Participation Overview}
Table~\ref{tab:team_participation} summarizes the participation of each team in the SCIDOCA 2025 Shared Task.
Out of seven registered teams, three—LA, VN25, and ZLB—submitted systems, covering different subsets of the three tasks.

\begin{table}[h]
    \centering
    \begin{tabular}{|c|c|c|c|}
        \hline
        \textbf{Team} & \textbf{Task 1} & \textbf{Task 2} & \textbf{Task 3} \\
        \hline
        LA   & \ding{51} & --         & --         \\
        VN25 & \ding{51} & \ding{51}  & \ding{51}  \\
        ZLB  & \ding{51} & \ding{51}  & \ding{51}  \\
        \hline
    \end{tabular}
    \caption{Team Participation Across Tasks}
    \label{tab:team_participation}
\end{table}

\subsection{Team LA}
This is the submission for the Task 1 of SCIDOCA shared tasks of team LA and the summary of our method is described as below:
In our citation prediction method, we begin by addressing the common discrepancy between the small number of correct citations within a text and the large pool of candidate abstracts. Instead of presenting an LLM with every possible abstract—which could lead to excessive computation and noise—we employ a comprehensive filtering mechanism. We use three distinct filters on the candidate documents: TF-IDF, dense retrieval, and a relation-based filter. First, the TF-IDF filter selects documents whose lexical terms significantly overlap with the query text, ensuring that we retain only those abstracts sharing relevant keywords or keyphrases. Next, we incorporate dense retrieval, which extends beyond exact term matching to capture semantic similarities, thus focusing on abstracts that demonstrate deeper contextual alignment with the query. Lastly, our relation-based filter extracts important relational information from the query text using an LLM, identifying relationships among entities, concepts, or themes within the text. We then compare these extracted relations to the candidate abstracts, selecting the top-k documents that share the most similar relational structure or content.

Once the candidate set is narrowed down by these three filters, we construct a specialized prompt by combining the query text and each remaining abstract. This prompt is submitted to the LLM, which is asked to decide whether the abstract should be cited and to provide a rationale for that decision. In doing so, we gather both a prediction—citable or not—and a chain of reasoning that explains why the model arrived at that conclusion. Rather than supplying this raw output directly, we further refine it by issuing a second prompt to the LLM that simplifies the chain of thought and consolidates its key points. This step helps maintain clarity and reduces extraneous details. Finally, we extract the predicted citations from the refined explanation. By uniting thorough filtering steps with iterative LLM prompting, our method efficiently pinpoints the abstracts most relevant to the given text and illuminates the reasoning behind each citation decision, thereby enhancing both accuracy and interpretability in citation prediction.

\subsection{Team VN25}
Vectorization is a critical step for transforming textual data into machine-readable representations,
supporting tasks such as citation discovery and retrieval. Traditional techniques like TF-IDF and word
embeddings (e.g., Word2Vec, GloVe) can encode fundamental semantics but often struggle with
long-range contextual dependencies typical of academic texts. More advanced transformer-based
models (e.g., BERT, RoBERTa) leverage self-attention to capture nuanced language patterns, yet their
512-token input limit poses challenges in processing extended passages where crucial context spans
multiple sentences.
Longformer employs a hybrid attention mechanism—combining local windowed attention and sparse
global attention—to process sequences of up to 4096 tokens. Unlike methods that rely on segmenting
inputs into smaller chunks or sliding windows, Longformer can handle most academic paragraphs in a
single pass, preserving broader contextual integrity. This capability is particularly beneficial for
citation-related tasks, as it ensures that citations and their surrounding context remain intact within one
continuous representation.
By adopting Longformer for vectorization, our approach maximizes context retention while keeping
computational overhead manageable. This design aligns well with prior research on extended
transformers, as it reinforces the importance of a more comprehensive encoding strategy for scholarly
texts. Consequently, we achieve greater accuracy and reliability in extracting citation information,
reflecting the growing trend toward leveraging longer sequence models for document-level
understanding.

The model is a binary classification system designed to predict whether a given text can cite a candidate.
Each candidate has a title and an abstract, and each of these attributes attends to multiple other texts.
These texts have already cited the candidate (this information is taken from the training set). As a result,
the candidate generates two context vectors—one for the title and one for the abstract. These vectors
are then concatenated with the given text and passed through a fully connected layer for final
classification.
In Task 1, the given text is the paragraph, and each candidate is one of the references in the
"candidate\_references" list.

In Task 2, for each [MASK] in the paragraph, we create a separate given text where only one specific
mask is kept, and all other masks are removed.
For Task 3: Citation Sentence Prediction, each sentence in the paragraph becomes a separate given text,
and each given text is paired with each candidate reference to create individual samples.

\subsection{Team ZLB}
ZLB focusing on Citation Discovery (Task 1) and Citation Sentence Prediction (Task 3). We propose an information retrieval approach utilizing transformer models to assess semantic similarity between query passages/sentences and candidate citations, which are represented by the concatenation of their titles and abstracts. Experiments were conducted using bge-large-en-v1.5, gte-qwen2-1.5b-instruct, and bge-reranker-large. Results demonstrate that bge-large-en-v1.5 outperforms gte-qwen2-1.5b-instruct in F1 score for Task 1, and fine-tuning bge-large-en-v1.5 achieved F1 score at 45\% . Notably, bge-reranker-large significantly outperforms bge-large-en-v1.5 in Task 3, achieving F1 score of 66\%, highlighting the effectiveness of cross-encoder architectures for sentence-level citation prediction.

\section{Result Analysis and Discussions}

\subsection{Task 1: Citation Discovery}

\textbf{Objective:} Predict relevant citations for a paragraph without specifying the exact sentence where the citation belongs.

\begin{table}[h]
    \centering
    \begin{tabular}{|c|r|r|r|}
        \hline
        \textbf{Team} & \textbf{Precision} & \textbf{Recall} & \textbf{F1-score} \\
        \hline
        LA   & 42.55  & \textbf{81.65}  & \textbf{50.75}  \\
        ZLB  & \textbf{49.26}  & 35.92  & 36.87  \\
        VN25 & 28.90  & 15.54  & 16.68  \\
        \hline
    \end{tabular}
    \caption{Performance on Task 1 (Ranked by F1-Score)}
    \label{tab:task1}
\end{table}

Team LA achieved the highest F1 score of 50.75 with a notably high recall of 81.65, indicating strong coverage of ground-truth citations. However, its moderate precision of 42.55 suggests some over-prediction. ZLB demonstrated a more balanced tradeoff between precision and recall but ultimately scored a lower F1 (36.87). VN25 had the lowest performance with limited recall and precision.

\subsection{Task 2: Masked Citation Prediction}

\textbf{Objective:} Predict the correct citation for each masked citation slot within a paragraph where the citation has been removed.

\begin{table}[h]
    \centering
    \begin{tabular}{|c|r|r|r|}
        \hline
        \textbf{Team} & \textbf{Precision} & \textbf{Recall} & \textbf{F1-score} \\
        \hline
        ZLB  & \textbf{28.86} & \textbf{28.86} & \textbf{28.86} \\
        VN25 &  6.60 &  6.60 &  6.60 \\
        LA   & -     & -     & -     \\
        \hline
    \end{tabular}
    \caption{Performance on Task 2 (Ranked by F1-Score)}
    \label{tab:task2}
\end{table}

Only ZLB and VN25 participated in Task 2. ZLB performed best with an F1 score of 28.86, demonstrating a strong capability in identifying the correct citations for masked slots. VN25 lagged behind significantly with an F1 score of just 6.60.

\subsection{Task 3: Citation Sentence Prediction}

\textbf{Objective:} Given a paragraph, predict the correct citation for each sentence that contains a citation.

\begin{table}[h]
    \centering
    \begin{tabular}{|c|r|r|r|r|}
        \hline
        \textbf{Team} & \textbf{Precision} & \textbf{Recall} & \textbf{F1-score} & \textbf{Accuracy (No Citation)} \\
        \hline
        ZLB  & \textbf{21.13} & \textbf{34.24} & \textbf{23.27} & \textbf{67.00} \\
        VN25 &  7.93 & 21.62 & 10.40 & 49.21 \\
        LA   & -     & -     & -     & -     \\
        \hline
    \end{tabular}
    \caption{Performance on Task 3 (Ranked by F1-Score)}
    \label{tab:task3}
\end{table}

ZLB again led in Task 3 with the highest F1 score (23.27) and the highest "Accuracy (No Citation)" of 67.00, indicating effective differentiation between citation and non-citation sentences. VN25 showed limited performance with an F1 of 10.40 and lower accuracy.

\subsection{Discussion}

The results highlight the varying strengths of participating teams across different citation tasks. Team LA achieved the best performance in Task 1, emphasizing coverage, whereas ZLB maintained solid performance across all three tasks, particularly excelling in structured prediction (Tasks 2 and 3). VN25, despite participating in all tasks, consistently scored lower, indicating room for improvement in citation-aware modeling and context understanding.

These initial results demonstrate the complexity of citation-related NLP tasks and set a benchmark for future systems to improve upon.


\section{Conclusion}
The SCIDOCA 2025 Shared Task aimed to advance research in citation discovery and prediction through three carefully designed subtasks. 
By releasing a high-quality, large-scale dataset derived from S2ORC and defining clear evaluation protocols, we provided a standardized benchmark for citation modeling systems.
Participation from diverse teams demonstrated a range of approaches and highlighted the challenges inherent in each task, particularly in balancing precision and recall in citation prediction.
The results show promising progress, especially in the citation discovery task, while also revealing significant room for improvement in more fine-grained prediction settings. 
We hope that the shared task outcomes, along with the released datasets and evaluation tools, will inspire continued research and development in automatic citation analysis and support the broader goal of improving scholarly information access and understanding.


\begin{credits}
\subsubsection{\ackname} This work was supported by "Strategic Research Projects" grant from ROIS (Research Organization of Information and Systems), Japan and JSPS KAKENHI Grant Number JP23K16954. Any opinions, findings, and conclusions or recommendations expressed in this material are those of the author(s) and do not necessarily reflect the views of the author(s)’ organization, JSPS or MEXT.
\end{credits}

%
%
%
\bibliographystyle{plain}
\bibliography{ref}
\end{document}